\documentclass[%
 reprint,
superscriptaddress,
 amsmath,amssymb,
 aps,
prb,
]{revtex4-2}

\usepackage{graphicx}
\usepackage{dcolumn}
\usepackage{bm}
\usepackage{hyperref}
\usepackage{soul}

\usepackage{afterpage}
\usepackage{xcolor}
\hypersetup{
    colorlinks=true,
    linkcolor={red!50!black},
    citecolor={blue!50!black},
    urlcolor={blue!80!black}
}

\begin{document}

\preprint{APS/123-QED}

\title{Dimensional crossover in spin Hall oscillators}

\author{Andrew Smith}
\affiliation{Department of Physics and Astronomy, University of California, Irvine, CA 92697, USA}
\author{Kemal Sobotkiewich}
\affiliation{Department of Physics, Center for Complex Quantum Systems, The University of Texas at Austin, Austin, Texas 78712, USA}
\author{Amanatullah Khan}
\affiliation{Department of Physics and Astronomy, University of California, Irvine, CA 92697, USA}
\author{Eric A. Montoya}
\affiliation{Department of Physics and Astronomy, University of California, Irvine, CA 92697, USA}
\author{Liu Yang}
\affiliation{Department of Physics and Astronomy, University of California, Irvine, CA 92697, USA}
\author{Zheng Duan}
\affiliation{Department of Physics and Astronomy, University of California, Irvine, CA 92697, USA}
\author{Tobias Schneider}
\affiliation{Helmholtz-Zentrum Dresden--Rossendorf, Institute of Ion Beam Physics and Materials Research, Bautzner Landstrasse 400, 01328 Dresden, Germany}
\author{Kilian Lenz}
\affiliation{Helmholtz-Zentrum Dresden--Rossendorf, Institute of Ion Beam Physics and Materials Research, Bautzner Landstrasse 400, 01328 Dresden, Germany}
\author{J{\"u}rgen Lindner}
\affiliation{Helmholtz-Zentrum Dresden--Rossendorf, Institute of Ion Beam Physics and Materials Research, Bautzner Landstrasse 400, 01328 Dresden, Germany}
\author{Kyongmo An}
\affiliation{Department of Physics, Center for Complex Quantum Systems, The University of Texas at Austin, Austin, Texas 78712, USA}
\author{Xiaoqin Li}
\affiliation{Department of Physics, Center for Complex Quantum Systems, The University of Texas at Austin, Austin, Texas 78712, USA}
\author{Ilya N. Krivorotov}
\affiliation{Department of Physics and Astronomy, University of California, Irvine, CA 92697, USA}

\date{\today}

\begin{abstract}
Auto-oscillations of magnetization driven by direct spin current have been previously observed in multiple quasi-zero-dimensional (0D) ferromagnetic systems such as nanomagnets and nanocontacts. Recently, it was shown that pure spin Hall current can excite coherent auto-oscillatory dynamics in quasi-one-dimensional (1D) ferromagnetic nanowires but not in quasi-two-dimensional (2D) ferromagnetic films. Here we study the 1D to 2D dimensional crossover of current-driven magnetization dynamics in wire-based Pt/$\mathrm{Ni}_{80}\mathrm{Fe}_{20}$ bilayer spin Hall oscillators via varying the wire width.  We find that increasing the wire width results in an increase of the number of excited auto-oscillatory modes accompanied by a decrease of the amplitude and coherence of each mode. We also observe a crossover from a hard to a soft onset of the auto-oscillations with increasing the wire width. The amplitude of auto-oscillations rapidly decreases with increasing temperature suggesting that interactions of the phase-coherent auto-oscillatory modes with incoherent thermal magnons plays an important role in suppression of the auto-oscillatory dynamics. Our measurements set the upper limit on the dimensions of an individual spin Hall oscillator and elucidate the mechanisms leading to suppression of coherent auto-oscillations with increasing oscillator size.


\end{abstract}

\pacs{Valid PACS appear here}
\maketitle

\begin{figure*}
\includegraphics[width=\textwidth]{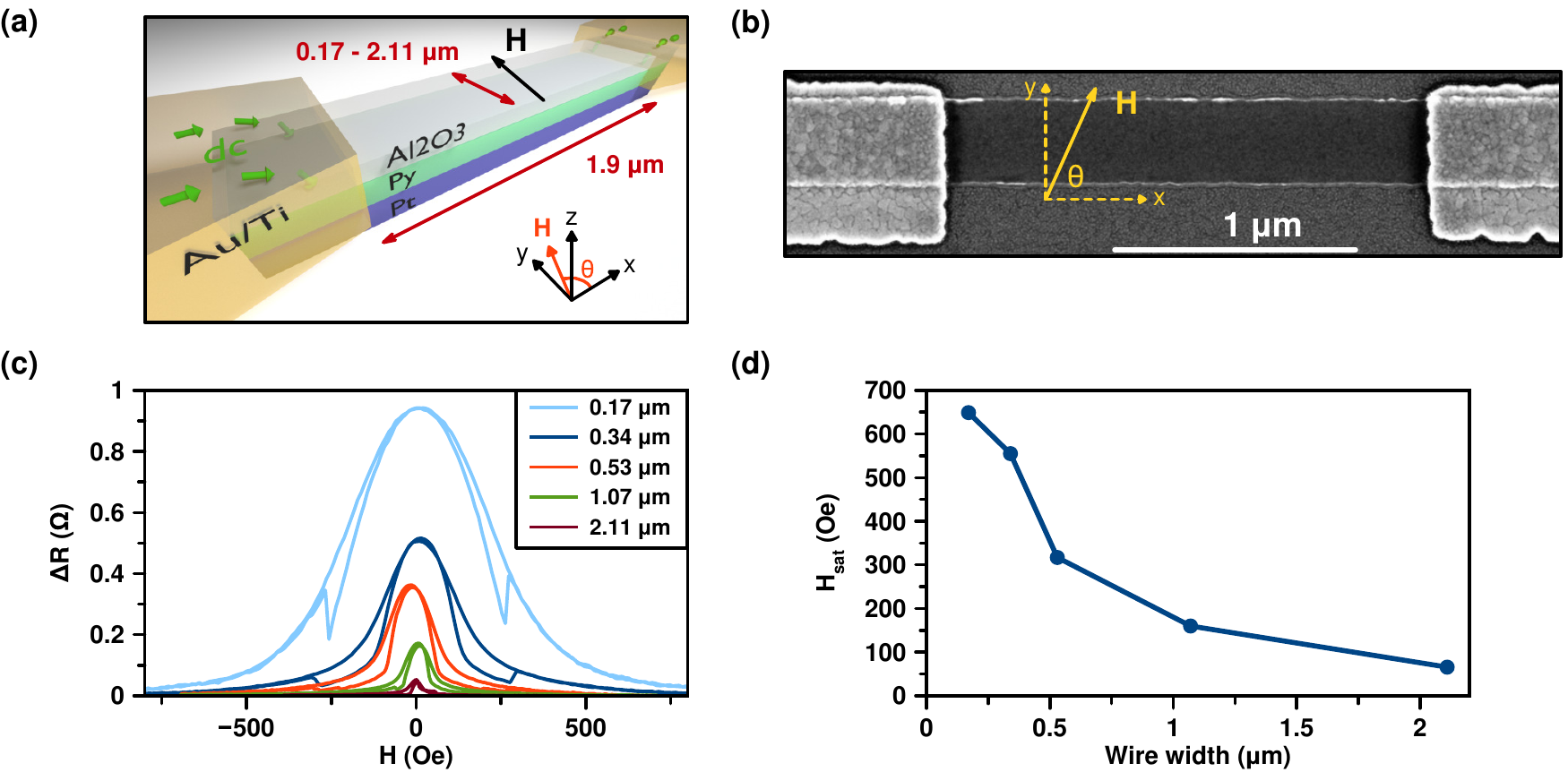}
\caption{\label{fig:Schematic_and_MR}
(a) Schematic of Pt/Py/AlO$_x$ wire SHO device. Five devices with the wire width varying from 0.17 $\mu$m to 2.11 $\mu$m were studied. The active region length is 1.9 $\mu$m for all devices.  Green arrows schematically show the electric current flow direction. (b) Scanning electron micrograph of the 0.51 $\mu$m wide SHO device. The magnetic field is applied in the plane of the sample at an angle $\theta$ with respect to the charge current flow direction. (c) Change in resistance of the five samples with different wire widths measured at 4.2 K for magnetic field applied at 5$^{\circ}$ off the $y-$axis ($\theta = 85^{\circ}$). (d) Saturation field $H_\mathrm{sat}$ of the wire for magnetic field applied at $\theta = 85^{\circ}$ as a function of the wire width (solid line is guide to the eye).}
\end{figure*}

\section{\label{sec:intro}Introduction} 

A spin current injected into a ferromagnet applies spin torque (ST) to its magnetization and can act as effective negative magnetic damping \cite{Slonczewski1996,Berger1996}. At a critical spin current density, the effective negative damping from spin torque overcomes the positive natural magnetic damping of the ferromagnet, which can result in the excitation of persistent phase-coherent auto-oscillatory dynamics of magnetization in the GHz frequency range \cite{Kiselev2003, Rippard2004, Krivorotov2005, Bertotti2007, Slavin2009, Deac2008, Collet2016, Awad2017}. Devices based on such auto-oscillations called spin torque oscillators (STOs) \cite{Chen2016} can be used as tunable sources of microwave radiation \cite{Tulapurkar2005,Safranski2019}, spin wave generators and amplifiers for nanomagnonic applications \cite{Madami2011,Urazhdin2014,Giordano2014,An2014}, sensors and magnetic field amplifiers in magnetic recording \cite{Braganca2010} and core building blocks of artificial neural networks for neuromorphic computing, image processing and pattern recognition \cite{Locatelli2014, Pufall2015, Torrejon2017}. 

Most STOs studied up to date are effectively zero-dimensional (0D) because the spin current in these devices is applied to a nanometer-scale region of the ferromagnet. Examples of such oscillators include (i) nanopillar spin valves \cite{Houssameddine2007, Pribiag2007, Bonetti2009,Villard2010,Rowlands2012, Choi2016} and nanoscale magnetic tunnel junctions \cite{Tulapurkar2005, Dussaux2010, Zeng2012, Rowlands2013, Skowronski2012}, (ii) nanocontacts to magnetic films and multilayers \cite{Tsoi2000, Mancoff2005, Silva2008,  Mistral2008, Mohseni2011a, Mohseni2013a, Bonetti2015, Demidov2012} and planar nanoconstriction spin Hall oscillators (SHOs) \cite{Liu2012a, Tarequzzaman2019, Spicer2018}. It was recently demonstrated that effectively one-dimensional (1D) SHOs can be realized in ferromagnetic nanowires where several low frequency spin wave eigenmodes are simultaneously driven into large-amplitude auto-oscillations by pure spin Hall current uniformly applied to a micrometer-scale active region of the nanowire \cite{Duan2014, Yang2015, Evelt2018, Wagner2018}. In contrast to the 0D and 1D geometries, STOs with effectively two-dimensional (2D) active region have not been realized. Brillouin light scattering measurements \cite{Demidov2011, Montoya2015} demonstrated that spatially uniform injection of spin current into an extended ferromagnetic film does not result in excitation of large-amplitude coherent auto-oscillations. Such suppression of auto-oscillatory dynamics in the 2D case was attributed to nonlinear coupling among the continuum of spin wave modes in the spatially extended film geometry \cite{Demidov2011}. As a result of such coupling, an increase of the amplitude of any spin wave mode leads to enhanced scattering of this mode into the continuum of other spin wave modes, which prevents any single mode from entering the regime of large-amplitude auto-oscillations.  

In this paper, we experimentally investigate the crossover between the 1D and 2D SHO geometries in order to elucidate the mechanisms responsible for suppression of coherent auto-oscillatory dynamics in the 2D limit. We study the current-driven  \cite{Tserkovnyak2006,Miron2011, Freimuth2014, Loconte2014, Amin2016, SeungHam2017, Humphries2017, Ado2017, Manchon2018, Baek2018, DC2018} dynamics in SHOs made from bilayers of Pt and Permalloy (Py = Ni$_{80}$Fe$_{20}$) where the anti-damping spin Hall current \cite{Zhang2000, Ando2008, Hoffmann2013, Liu2014, Mendes2014, Tshitoyan2015, Nan2015, Zhang2017, Takeuchi2018, Sagasta2018, Solyom2018} generated in Pt is injected into the Py layer. The 1D to 2D crossover is realized via varying the wire width from the nanometer- to the micrometer-scale dimensions.  Quite surprisingly, we find that there is an optimal wire width that maximizes the amplitude and coherence of the auto-oscillatory dynamics. Our measurements reveal that the number of auto-oscillatory modes increases with increasing the wire width beyond the optimal, while their amplitude and phase coherence \cite{Iacocca2014} both decrease. A similar result was found in a recent independent unpublished study \cite{Lee2020}. We find that the transition between the large-amplitude coherent auto-oscillatory dynamics in the 1D geometry and the small-amplitude incoherent dynamics in the 2D geometry is a continuous crossover.  We also report measurements of the auto-oscillatory dynamics as a function of temperature, and observe a rapid decrease of the amplitude of auto-oscillations with increasing temperature, which points to the important role of  the auto-oscillatory mode scattering on thermal magnons in limiting the amplitude and coherence of the auto-oscillations. 

\section{\label{sec:techniques}Experimental techniques} 

\subsection{\label{sec:exp}Samples and magnetoresistance}

The Pt/Py wire samples are patterned using e-beam lithography and ion milling from a (c-plane sapphire substrate)/Pt(6 nm)/Py(5 nm)/AlO$_x$(2 nm) multilayer grown by magnetron sputtering. The wires of five different widths $w$: 0.17 $\mu$m, 0.35 $\mu$m, 0.53 $\mu$m, 1.07 $\mu$m and 2.11 $\mu$m and the length of 40 $\mu$m are studied. Two Ti(10 nm)/Au(40 nm) leads for application of electric current bias to the middle section of the wire are patterned atop the wire as shown in Figures \ref{fig:Schematic_and_MR}(a) and \ref{fig:Schematic_and_MR}(b). The 1.9 $\mu$m long middle section of the Pt/Py wire between the leads defines the active region of the wire SHO, where the electric current and the spin Hall current densities are much higher than those under the leads.

Due to the spin Hall effect \cite{Dyakonov1971, Sinova2015}, the electric current in the Pt layer generates a transverse spin current that flows perpendicular to the sample plane and is injected into the Py layer. The spin polarization vector of this current is in the sample plane perpendicular to the wire axis ($\theta=90^{\circ}$ as shown in Fig.~\ref{fig:Schematic_and_MR}(b)). In our measurements, the wire is magnetized in the plane of the sample by a magnetic field applied at an angle $\theta$ with respect to the wire axis as illustrated in Fig.~\ref{fig:Schematic_and_MR}(b). ST from the spin Hall current applied to the Py magnetization can act as negative magnetic damping. Specifically, the component of the spin current polarization parallel to the magnetization gives rise to the antidamping ST \cite{Ralph2008, Schulz2017, Belashchenko2019, Ryu2019}, which results in the maximum antidamping action of ST at $\theta=90^{\circ}$. When the antidamping ST exceeds the natural magnetic damping of Py \cite{Heinrich1967, Heinrich2005b}, it can excite auto-oscillatory dynamics of the Py magnetization \cite{Kiselev2003, Liu2012a, Sato2019}.

\begin{figure*}[htbp]
\includegraphics{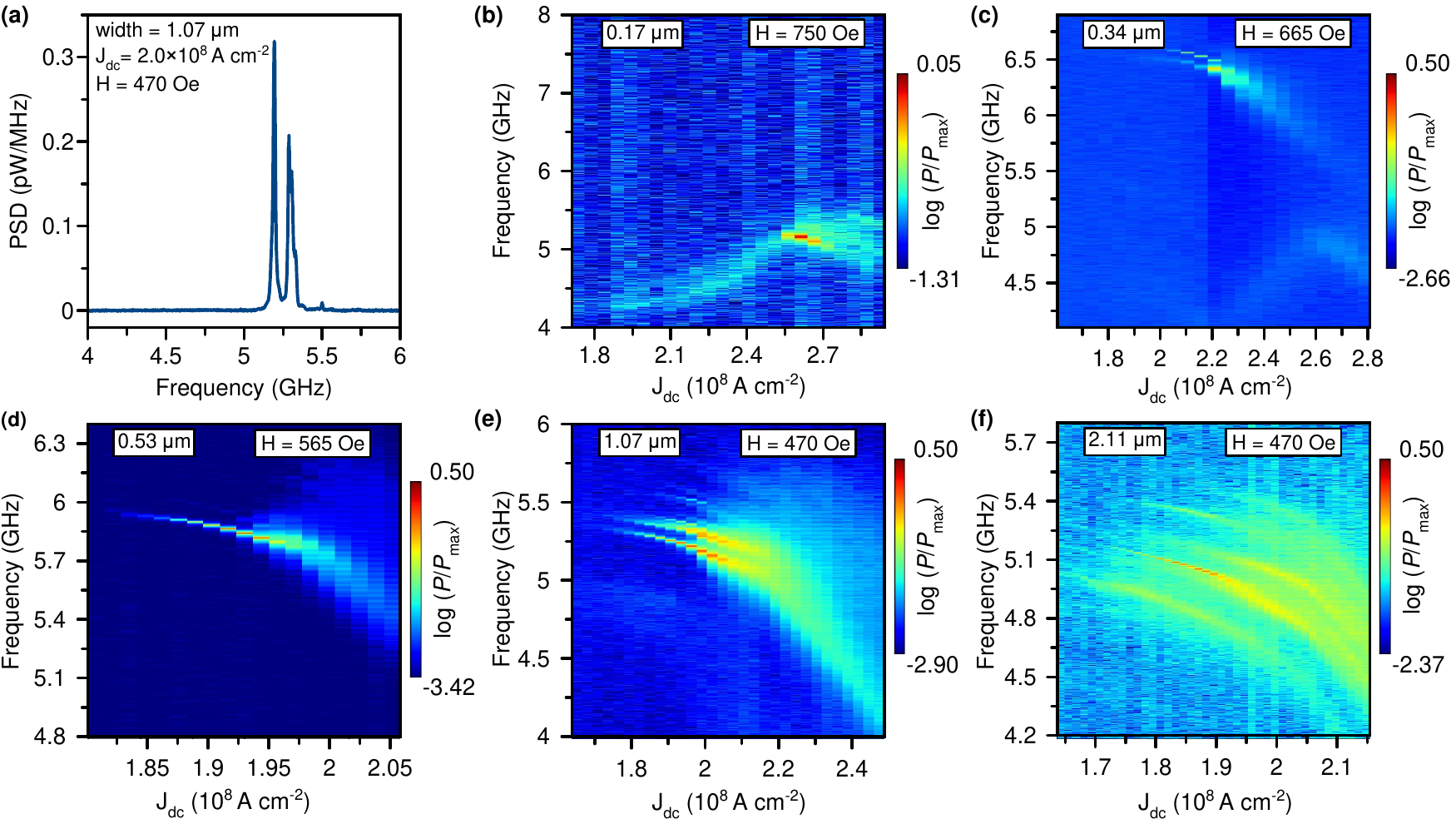}
\caption{\label{fig:Emission}
(a) Power spectral density (PSD) of the microwave signal generated by the 1.07 $\mu$m wide Pt/Py/AlO$_\mathrm{x}$ wire SHO at the bias current density $J_\mathrm{dc}=2.0\times 10^8$ A\,cm$^{-2}$, bath temperature $T$ = 4.2 K and magnetic field $H =$ 470 Oe applied at $5^{\circ}$ from $y-$axis $\left( \theta = 85^{\circ} \right)$. Three auto-oscillatory modes are excited. The non-Lorentzian peak lineshapes are mainly due to standing waves in the microwave circuit. (b--f) Microwave emission spectra versus $J_\mathrm{dc}$ for five wires of different widths measured in fields exceeding $H_\mathrm{sat}$ by approximately 100 Oe and applied at $\theta = 85^{\circ}$. The logarithmic color scale represents the emitted power normalized to the maximum power $P_\mathrm{max}$.  The wire widths and the applied field values are shown in the figures. The low frequency edge mode is seen for narrower wires (0.17 $\mu$m and 0.34 $\mu$m wide). The higher frequency bulk modes are observed in wider wires (0.34 $\mu$m, 0.53 $\mu$m, 1.07 $\mu$m and 2.11 $\mu$m wide).}
\end{figure*}

Fig.\,\ref{fig:Schematic_and_MR}(c) shows the field dependent change in resistance $\Delta R$ of five samples, differing from each other by the wire width, measured as a function of magnetic field $H$ applied in the sample plane nearly perpendicular to the wire axis ($\theta=85^{\circ}$) at a small bias current and a bath temperature of $T =$ 4.2 K. The observed field dependence of resistance $\Delta R(H)$ arises from anisotropic magnetoresistance (AMR) \cite{Mcguire1975, Kokado2013a} of Py, for which the resistance of the sample is maximum $R_\mathrm{max}$ when the magnetization is parallel to the electric current direction and is minimum $R_\mathrm{min}$ when magnetization is perpendicular to the electric current. The $5^{\circ}$ offset of applied magnetic field from $\theta=90^{\circ}$ is used to produce a measureable microwave emission signal, as described in the next section, due to the angular dependence of AMR. The magnitude of the AMR signal $\Delta R_\mathrm{AMR} = R_\mathrm{max} - R_\mathrm{min}$ decreases with increasing width of the wire because of the decreasing wire resistance $R$. The in-plane shape anisotropy field $H_\mathrm{a}$ of the Py wire is approximately equal to the AMR saturation field $H_\mathrm{sat}$ in Fig.~\ref{fig:Schematic_and_MR}(c). Fig.~\ref{fig:Schematic_and_MR}(d) shows the wire width dependence of $H_\mathrm{sat}$ defined as the field for which $R = R_\mathrm{min}+0.05\,\Delta R_\mathrm{AMR}$.  The saturation field $H_\mathrm{sat}$ increases with decreasing wire width due to the enhanced demagnetization field in narrower wires \cite{Aharoni1998}. We note that in the transversely magnetized nanowire geometry, the demagnetizing field is spatially non-uniform and is much larger near the wire edges compared to the wire center. For this reason, magnetization in the wire center saturates in the direction of the applied field at significantly smaller applied field (bulk saturation field) compared to the field required to saturate magnetization at the wire edge (edge saturation field) \cite{McMichael2006, Duan2015}. Since the characteristic length scale of the edge demagnetizing field localization is similar to the wire thickness \cite{Duan2015}, and the width of all our wires significantly exceeds their thickness, magnetization of a large fraction of our wire saturates at a field close to the bulk saturation field.  For this reason, $H_\mathrm{sat}$ defined above is similar to the bulk saturation field.

\subsection{\label{sec:emission}Microwave emission measurements}

To study the auto-oscillatory magnetic dynamics excited by spin Hall current, we saturate magnetization of the wire in the plane of the sample by a magnetic field $H > H_\mathrm{sat}$, apply a direct electric current $I_\mathrm{dc}$ to the wire and measure the microwave signal generated by the device using a spectrum analyzer \cite{Kiselev2003}. The microwave voltage $V_\mathrm{ac} \sim I_\mathrm{dc} \delta R_\mathrm{ac}$ is generated by the AMR resistance oscillations $\delta R_\mathrm{ac}$ arising from the magnetization auto-oscillations of the Py layer \cite{Liu2013}. 

Fig.\,\ref{fig:Emission} illustrates the spectral properties of the microwave signals generated by the Pt/Py wire devices as a function of direct electric current density $J_\mathrm{dc}$ applied to the wire at the sample bath temperature $T=\,$4.2 K and $\theta=85^{\circ}$. Here the charge current density $J_\mathrm{dc}$ is defined as direct current bias $I_\mathrm{dc}$ divided by the cross-sectional area of the Pt/Py bilayer wire. To facilitate direct comparison among the wires of different widths, the measurements in Fig.\,\ref{fig:Emission} are made at magnetic field values exceeding the width-dependent saturation field $H_\mathrm{sat}(w)$ by approximately 100 Oe. This ensures approximate spatial uniformity of the wire's saturation magnetization  and similarity of the internal magnetic fields among the wires of different widths \cite{Duan2015}.  Fig.~\ref{fig:Emission}(a) shows a typical microwave emission spectrum measured for the 1.07 $\mu$m wide wire device at  $J_\mathrm{dc}=2.0\times 10^8$ A\,cm$^{-2}$ and $H=$ 470 Oe. The spectrum shows multiple peaks, which demonstrates that auto-oscillations of several spin wave modes of the device are excited at this value of $J_\mathrm{dc}$. Such coexistence of multiple spin wave modes has been the focus of recent research \cite{Dumas2013, Zhang2017a}. Quantitative analysis of the spectral linewidth and lineshape of these peaks is complicated by the presence of standing waves in the microwave circuitry, which manifest themselves as oscillatory modulation of the spectral peak amplitude seen for the peak at 5.3 GHz.

\begin{figure*}[htbp]
\includegraphics{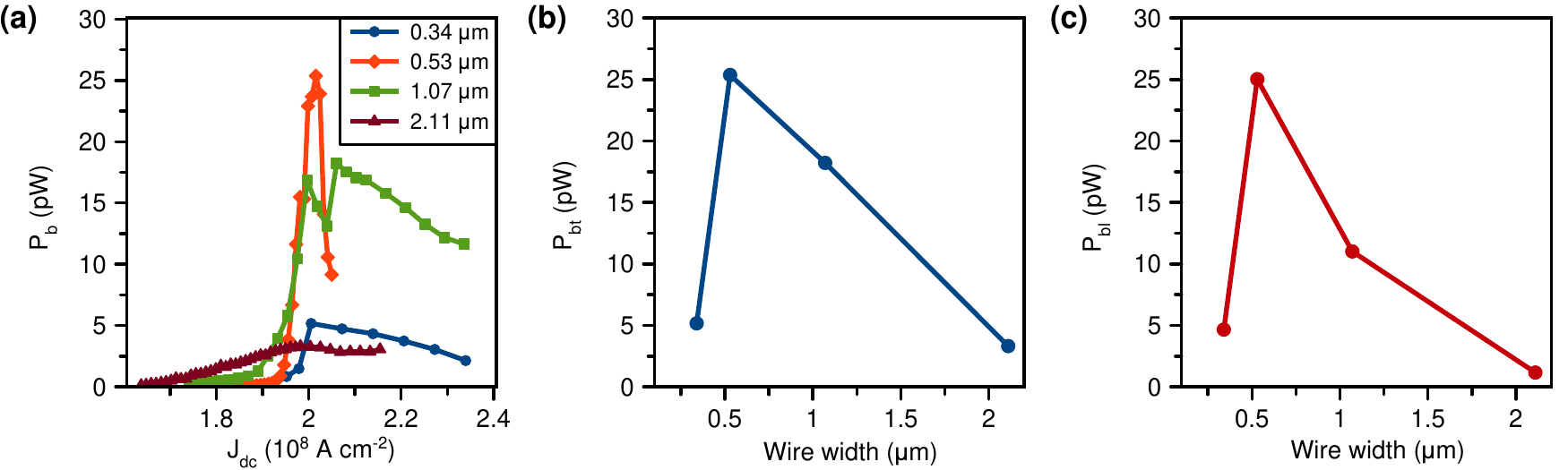}
\caption{\label{fig:Emission_Power}
(a) Integrated microwave power emitted by all bulk modes of the wire $P_\mathrm{b}$ versus bias current density $J_\mathrm{dc}$. The data is shown for all four wires exhibiting bulk mode auto-oscillations. (b) Maximum value of the integrated microwave power emitted by all bulk modes of the wire $P_\mathrm{bt}$ as a function of the wire width. (c)  Maximum integrated power of the largest-amplitude bulk mode $P_\mathrm{bl}$ as a function of the wire width.}
\end{figure*}

Figs.~\ref{fig:Emission}(b)--\ref{fig:Emission}(f) show the microwave emission spectra generated by five Pt/Py wire devices with different wire widths as a function of $J_\mathrm{dc}$. For the 0.17 $\mu$m wide wire (Fig.~\ref{fig:Emission}(b)), a single low-frequency mode is observed. The frequency of this mode first increases and then decreases with current \cite{Dvornik2018}, and the spectral line of this mode is relatively broad. Previous studies of the Pt/Py nanowire SHOs \cite{Duan2014} revealed that this low-frequency mode is the edge mode (EM) whose amplitude is maximum at the wire edge. This mode is created by a  spatially inhomogeneous demagnetizing field, which creates magnetic potential wells for spin waves near the edges of a transversely magnetized wire \cite{Duan2015, Park2002, Bayer2006, Duan2014a}. Since this mode is confined to a relatively small volume near the wire edges, random thermal torques are expected to significantly increase the spectral linewidth of this mode \cite{Slavin2009} compared to the bulk modes that occupy the entire active region.  The wire edge roughness and the magnetic material inhomogeneities caused by ion mill damage and oxidation of the edge may also contribute to the line broadening of the EM \cite{McMichael2006}. 

The EM auto-oscillations are also excited in the 0.34 $\mu$m wide wire (Fig.\,\ref{fig:Emission}(c)) but this mode generates less microwave power in comparison to the 0.17 $\mu$m wire.  This can be explained by the smaller volume fraction occupied by the EM in the 0.34 $\mu$m wire, and does not imply that the amplitude of the EM auto-oscillations in the wider wire is lower than that in the narrower wire. In addition to the EM, two higher frequency auto-oscillatory modes are excited in the 0.34 $\mu$m wire. Such high frequency modes with narrow spectral linewidths have been previously identified as bulk spin wave modes (BM), whose amplitudes are maximum near the center of the wire \cite{Duan2014, Duan2015}. The existence of multiple BM with a discrete set of eigenfrequencies \cite{Gui2007} is a result of geometric confinement of the magnetic oscillations to the SHO active region. The confinement along the wire width is provided by the geometric edges of the wire. The confinement along the wire length may arise from two sources: (i) a current-induced Oersted field that opposes the applied field for the current polarity generating anti-damping ST in the Pt/Py system \cite{Duan2014} and (ii) a step-wise change of the effective magnetic damping at the boundaries between the active region and the electric leads.  All observed BMs exhibit red frequency shifts with increasing current that arises from several factors: (i) reduction of the Py saturation magnetization via ohmic heating and short-wavelength magnon generation by ST \cite{Demidov2011}, (ii) Oersted field from the electric current in the Pt layer and (iii) nonlinear frequency shift due to increase of the mode amplitude with increasing current \cite{Slavin2009, Krivorotov2008, Boone2009}.

For the 0.53 $\mu$m wire (Fig.~\ref{fig:Emission}(d)), auto-oscillatory dynamics of the the EM are no longer detected. This, however, does not imply that the EM is not excited. As the volume fraction of the active region occupied by the EM becomes smaller, more of the applied current flowing in the ferromagnet shunts through the bulk of the wire and the microwave signal generated by magnetization oscillations at the wire edge falls below the detection threshold of our measurement setup. It is interesting to note that for this wire width, auto-oscillations of the lowest frequency BM are excited with large amplitude while the amplitude of the higher frequency BMs is negligibly small compared to that of the lowest frequency BM. At this wire width, the Pt/Py wire SHO behaves nearly as a single-mode microwave signal generator. As we discuss below, this single-mode behavior gives rise to the highest microwave power generated among all SHOs studied in this paper.

Further increase of the Pt/Py wire width results in excitation of the auto-oscillatory dynamics of multiple BMs. The number of the excited BM increases with increasing wire width, as illustrated in Fig.\,\ref{fig:Emission}(e) for the 1.07 $\mu$m wide wire and Fig.\,\ref{fig:Emission}(f) for the 2.11 $\mu$m wide wire, while the amplitude of auto-oscillations of each mode decreases. The precipitous decrease of the auto-oscillation amplitude with increasing wire width is illustrated in Fig.\,\ref{fig:Emission_Power}, which shows integrated microwave power generated by the auto-oscillatory modes. Fig.\,\ref{fig:Emission_Power}(a) shows the total integrated power in all BMs of a wire $ P_\mathrm{b} $ as a function of the applied current density $J_\mathrm{dc}$. This figure demonstrates that the critical current for the excitation of the auto-oscillatory dynamics  $J_\mathrm{c} \approx 2\times 10^8$ A\,cm$^{-2}$ is nearly independent on the wire width. For all wires, the integrated microwave power first increases with increasing current above $J_\mathrm{c}$ and then decreases after reaching a maximum at a current density $J_\mathrm{max}$. Fig.\,\ref{fig:Emission_Power}(a) also shows that the onset of the auto-oscillations becomes softer as the wire width increases. Indeed, the current interval between $J_\mathrm{c}$ and $J_\mathrm{max}$ is $\approx 0.5\times 10^7$ A\,cm$^{-2}$ for the 0.34 $\mu$m wide wire while it is $\approx 2.5\times 10^7$ A\,cm$^{-2}$ for the 2.11 $\mu$m wide wire.

We also find that the maximum integrated microwave power generated by the wire is a non-monotonic function of the wire width. Fig.\,\ref{fig:Emission_Power}(b) shows the maximum total integrated power generated by all BMs of a wire $P_\mathrm{bt}$ as a function of the wire width. The power first increases with the wire width reaching the value of 26 pW for the 0.53 $\mu$m wide wire and then rather precipitously decreases with increasing width. Comparison of this figure with Fig.\,\ref{fig:Emission} clearly shows that the maximum power is achieved in the 0.53 $\mu$m wide wire that exhibits single-mode auto-oscillatory dynamics. Fig.\,\ref{fig:Emission_Power}(c) illustrates that the power generated by the largest-amplitude BM $P_\mathrm{bl}$ shows a similar trend to that in Fig.\,\ref{fig:Emission_Power}(b). However the decrease of the largest-amplitude BM power with increasing wire width is even more rapid than that in Fig.\,\ref{fig:Emission_Power}(b). 

The data in Fig.\,\ref{fig:Emission} and Fig.\,\ref{fig:Emission_Power} clearly show that coherent auto-oscillatory dynamics of magnetization are rapidly suppressed as the wire width increases into the micrometer-scale range. The data demonstrate that the crossover between coherent large-amplitude auto-oscillatory dynamics in 1D nanowires and incoherent small-amplitude dynamics in 2D microwires proceeds via a gradual increase of the number of spin wave modes participating in the auto-oscillatory dynamics accompanied by a rapid decrease of the maximum amplitude of auto-oscillations achievable by each of these modes.

\subsection{\label{sec:BLS}Brillouin light scattering}

\begin{figure*}[!ht]
\includegraphics[width=\linewidth]{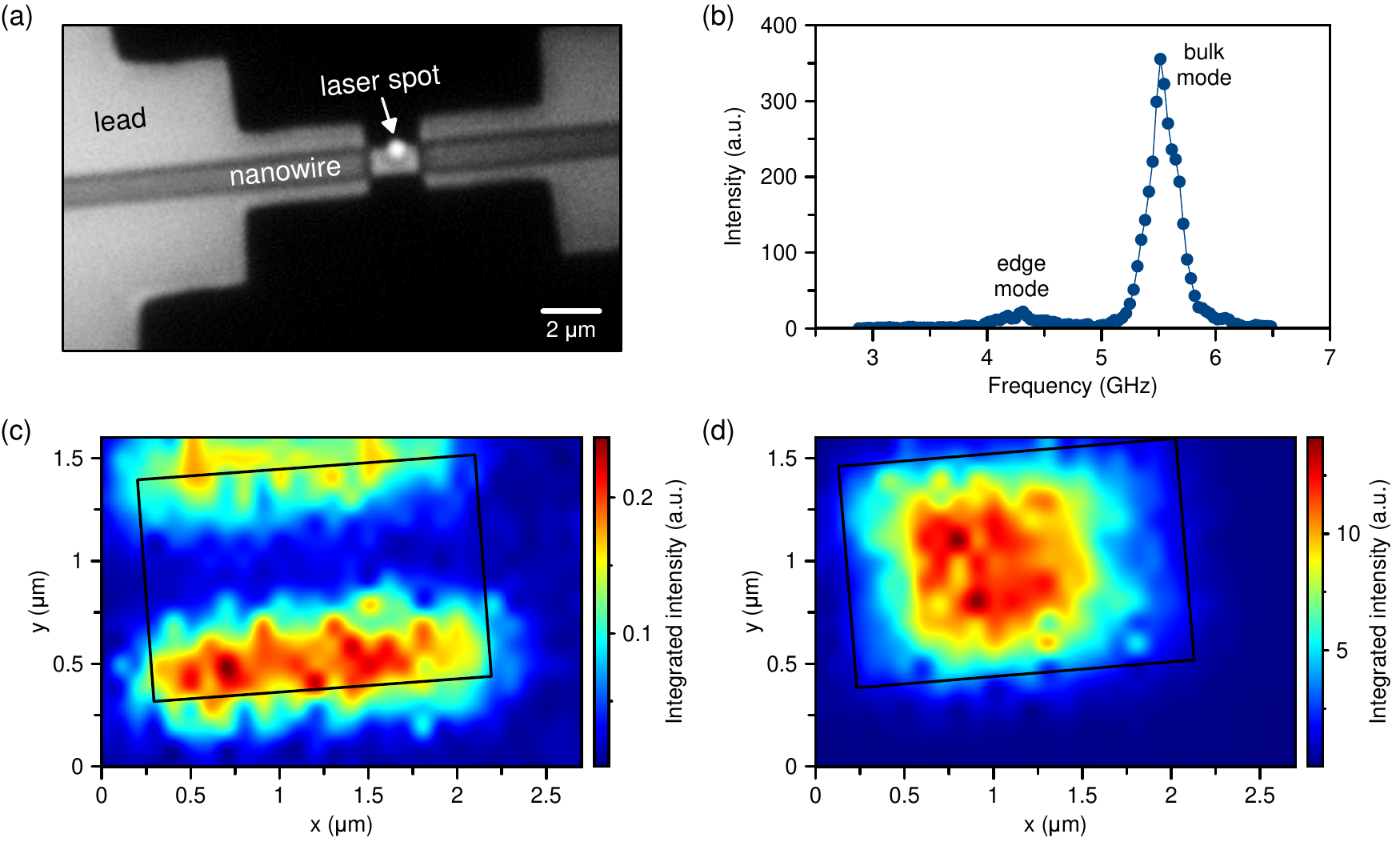}
\caption{\label{fig:BLS}
Brillouin light scattering characterization of auto-oscillatory modes of the 1.07 $\mu$m wide wire device at room temperature ($ T = 300\, \mathrm{K}$). (a) Optical micrograph of the devices showing Pt/Py wire, Ti/Au leads and laser spot of the BLS apparatus focused on one edge of the active region of the device.
(b) BLS spectrum measured at the laser beam position at the edge of the active region shown in (a) for $J_\mathrm{dc}=9.77\times 10^7$ A\,cm$^{-2}$ and $H=495$\,Oe applied at $\theta = 87^\circ$. Spatially resolved BLS allows to reveal the origin of the excitations:  
(c) Spatial profile of the BLS signal intensity at $f=4.3$\,GHz -- the center frequency of the low-frequency peak (edge mode). The rectangle denotes the approximate location of the active region of the device. This BLS spectral mapping reveals that this auto-oscillatory mode is an edge spin wave mode of the wire. (d) Spatial profile of the BLS signal intensity at $f=5.6$\,GHz -- the center frequency of the high-frequency peak (bulk mode). This BLS spectral mapping reveals that this auto-oscillatory mode is a bulk spin wave mode of the wire.}
\end{figure*}

We employ micro-Brillouin light scattering (BLS) \cite{Montoya2015} to directly measure the spatial profiles of the auto-oscillatory modes at room temperature ($ T = 300\, \mathrm{K}$). Fig. \ref{fig:BLS}(a) shows optical micrograph of the 1.07 $\mu$m wide wire device. The Pt/Py wire and the Ti/Au leads are marked in this image. The image also shows the laser beam of the BLS setup focused on one edge of the wire active region. Fig.~\ref{fig:BLS}(b) shows the BLS spectrum measured at this laser beam position. The frequency resolution of the BLS apparatus is $\sim 0.1~\textrm{GHz}$. The data are taken at a current density $J_\mathrm{dc}=9.77\times 10^7$ A\,cm$^{-2}$ exceeding the critical density for excitation of the auto-oscillatory dynamics at room temperature. The measurement is made in 495 Oe magnetic field applied in the plane of the sample nearly perpendicular to the wire axis ($\theta = 87^\circ$). Two peaks are visible in the spectrum. Spectral mapping of the BLS signal intensity at center frequencies of these peaks allows us to determine the spatial profiles of the excited auto-oscillatory modes. 

Fig.~\ref{fig:BLS}(c) shows the BLS signal intensity measured at the center frequency of the low-frequency peak $f=4.3$\,GHz as a function of the laser beam position. The rectangular frame on this figure outlines the contour of the active region of the device. It is clear from Fig.~\ref{fig:BLS}(c) that the low-frequency auto-oscillatory mode is the edge mode of the sample (labeled as edge mode in Fig.~\ref{fig:BLS}(b)). It is interesting to note that the amplitude of this mode at one edge exceeds that at the other edge. This symmetry breaking may result from different edge roughness of the two edges \cite{McMichael2006}.  Fig.~\ref{fig:BLS}(d) shows the BLS signal intensity measured at the center frequency of the high-frequency peak $f=5.6$\,GHz as a function of the laser beam position. This figure reveals that the high frequency mode is the bulk mode of the wire (labeled as bulk mode in Fig.~\ref{fig:BLS}(b)). 

\subsection{\label{sec:micromagnetic}Micromagnetic simulations}

\begin{figure}[htbp]
\includegraphics{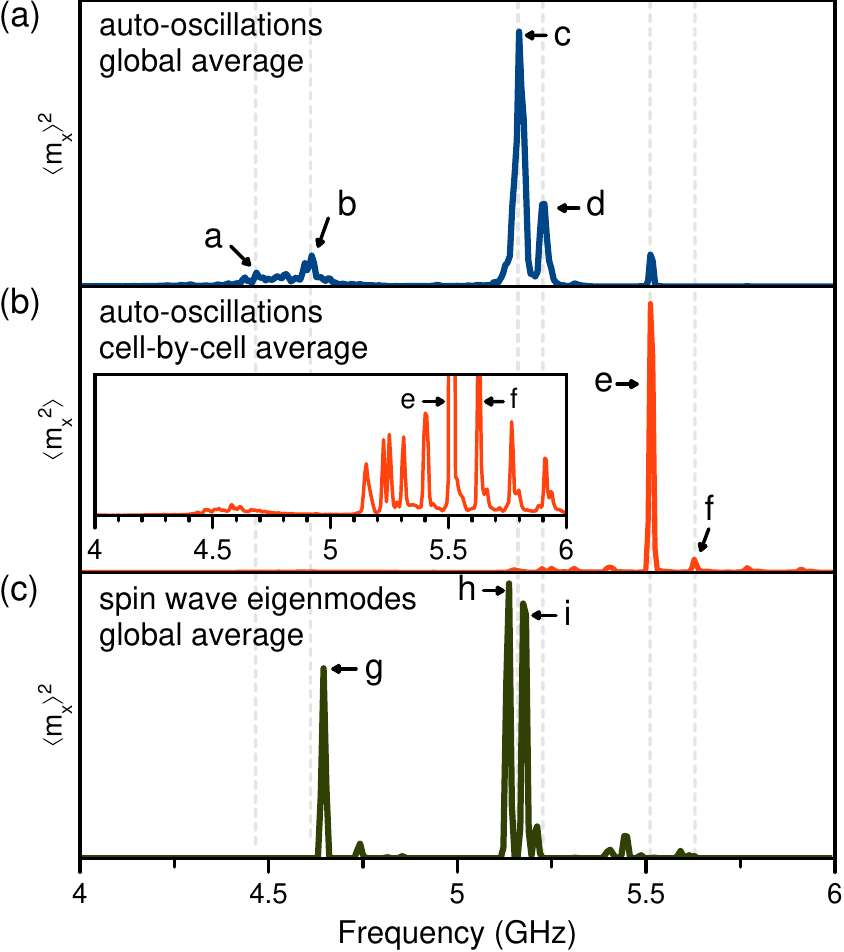}
\caption{\label{fig:Spectra}
Micromagnetic simulations of spin wave spectra in the 1.07 $\mu$m wide wire with $H_\mathrm{Oe} = 36$ Oe, $H$ = 470 Oe and $\theta = 85^\circ$. 
(a) Simulated spin Hall oscillator microwave emission spectra at $\theta_\mathrm{SH} = 0.045$ (global average FFT) (b) Simulated cell-by-cell FFT spectra $\langle m_x^2 \rangle$ at $\theta_\mathrm{SH} = 0.045$ to be compared to BLS data. Note the scale of (b) is $2500 \times$ the scale of (a).  Inset shows a zoom of $50 \times$ in amplitude. (c) Simulated spin wave eigenmode spectra (global average FFT). }
\end{figure}

In order to qualitatively understand the types of auto-oscillatory modes excited in our microwave emission and BLS experiments, we perform micromagnetic simulations of current-driven magnetization dynamics in the 1.07 $\mu$m wide wire at zero temperature ($ T = 0 \,\mathrm{K}$). The simulations are carried out using the Object Oriented Micromagnetic Framework (OOMMF) \cite{Donahue1999} using previously determined magnetic parameters of the Py film on Pt underlayer \cite{Duan2014, Duan2014a}: saturation magnetization $ M_\mathrm{s} = 620$ emu\,cm$^{-3}$, exchange constant $ A_\mathrm{ex} = 5\times 10^{-12}$ J\,m$^{-1}$. We assume a Gilbert damping constant $\alpha=0.01$. We micromagnetically simulate the current-driven auto-oscillatory magnetic dynamics via application of the anti-damping spin Hall torque to the active region of the Pt/Py wire device. We then record the spatial and time dependence of the normalized magnetization component along the wire axis $m_x \left( t \right)$ over the active region. We repeated these simulations for several initial  directions of magnetization and found the results to be independent of the initial conditions.

We simulate a wire of 4 $\mu$m total length with a 1.9 $\mu$m active region in the center. The Py layer is divided into $ 5 \times 5 \times 5 \, \mathrm{nm}^3 $ micromagnetic cells. In the active region, we apply an in-plane Oersted field $H_\mathrm{Oe} = 36$ Oe parallel the $y$-axis, which corresponds to a room-temperature critical current density in the Pt layer of $J_\mathrm{c}^{\mathrm{Pt}}= 9.5\times 10^7$ A\,cm$^{-2}$ \cite{Wagner2018}. This Oersted field opposes the external magnetic field and creates a magnetic potential well for spin waves, resulting in the localization of the auto-oscillations within the active region. In the active region, we also apply antidamping spin Hall torque with spin Hall angle $\theta_\mathrm{SH}$ parameterizing the conversion of charge current in Pt into transverse spin current injected from Pt into Py. We vary the spin Hall angle and observe the clear emergence of auto-oscillatory behavior at spin Hall angles exceeding the critical value $\theta_\mathrm{SH}^\mathrm{c} = 0.044$. We report simulation results just above this critical angle ($\theta_\mathrm{SH} = 0.045 > \theta_\mathrm{SH}^\mathrm{c}$) to analyze the auto-oscillatory behavior at a current density exceeding the critical current density by approximately 2$\,\%$. We find that transient dynamics lasts for approximately 0.1 $\mu$s after turning on the spin Hall current. After this period of time, quasi-steady state dynamics are achieved. In the quasi-steady state, the same set of auto-oscillatory modes are excited, however their amplitudes fluctuate as a function of simulation time. We attribute this to zero-temperature deterministic chaos which can arise in such nonlinear dynamical systems with a large number of degrees of freedom. The spectra of auto-oscillatory dynamics are calculated via fast Fourier transforms (FFT) of the $x$-component of the dynamic magnetization $m_x \left( t \right)$ in the active region. The FFT spectra and spatial profiles are calculated from simulation results using a 0.2 $\mu$s -- 2.0 $\mu$s time interval after the start of the simulations.

In order to compare the simulated spectra with the microwave emission and BLS experiments, we must consider that the two measurements are sensitive to $m_x \left( t \right)$ in different ways. For the electrically detected emission experiment, the spin Hall oscillator output power $P_\textrm{SHO}$ is proportional to the square of the dynamic magnetoresistance oscillations $\delta R_\mathrm{ac}$, which are approximately proportional to the average $m_x \left( t \right)$ in the active region \cite{Duan2014},
\begin{equation}\label{eq:P_SHO}
  P_\textrm{SHO} \propto  \delta R_\mathrm{ac}^2 \propto  \langle m_x \rangle ^2.
\end{equation}
Therefore to simulate the emission spectra, we first take the \textit{global average} of $m_x \left( t \right)$ in the active region and then take the square of the FFT spectra. Out-of-phase magnetization dynamics in different regions of SHO contribute destructively to $\delta R_\mathrm{ac}$, and thus  $P_\textrm{SHO}=0$ does not prove the absence of auto-oscillations. The BLS experiment, on the other hand, is sensitive to the total magnon population and  the BLS signal is proportional to the square of the dynamic magnetization. Therefore to emulate the BLS spectra, we calculate the square of the FFT spectra for each micromagnetic region \textit{cell-by-cell} and then take the average; in this manner we simulate the BLS peak intensity,
\begin{equation}\label{eq:I_BLS}
  I_\textrm{BLS} \propto  \langle m_x^2 \rangle.
\end{equation}
We note that Eq.(\ref{eq:I_BLS}) is only a qualitative estimate because the BLS peak intensity also depends on the wave vector of the spin wave and drops to zero above a critical value of the wave vector \cite{Sebastian2015}. 

Fig.~\ref{fig:Spectra} shows the micromagnetic simulation spectrum for the 1.07 $\mu$m wide wire in an external in-plane field of $H = 470~\textrm{Oe}$ applied 5$^{\circ}$ from the $y$-axis ($\theta = 85^\circ$). Fig.~\ref{fig:Spectra}(a) shows the global average spectrum simulating the microwave emission experiment and reveals that two clusters of large amplitude peaks, one just above 4.5 GHz and one just above 5.0 GHz, are simultaneously excited above the critical current. Fig.~\ref{fig:Spectra}(b) shows the cell-by-cell average spectrum to be compared to the BLS data.

\begin{figure*}[!ht]
\includegraphics[width=0.95\textwidth]{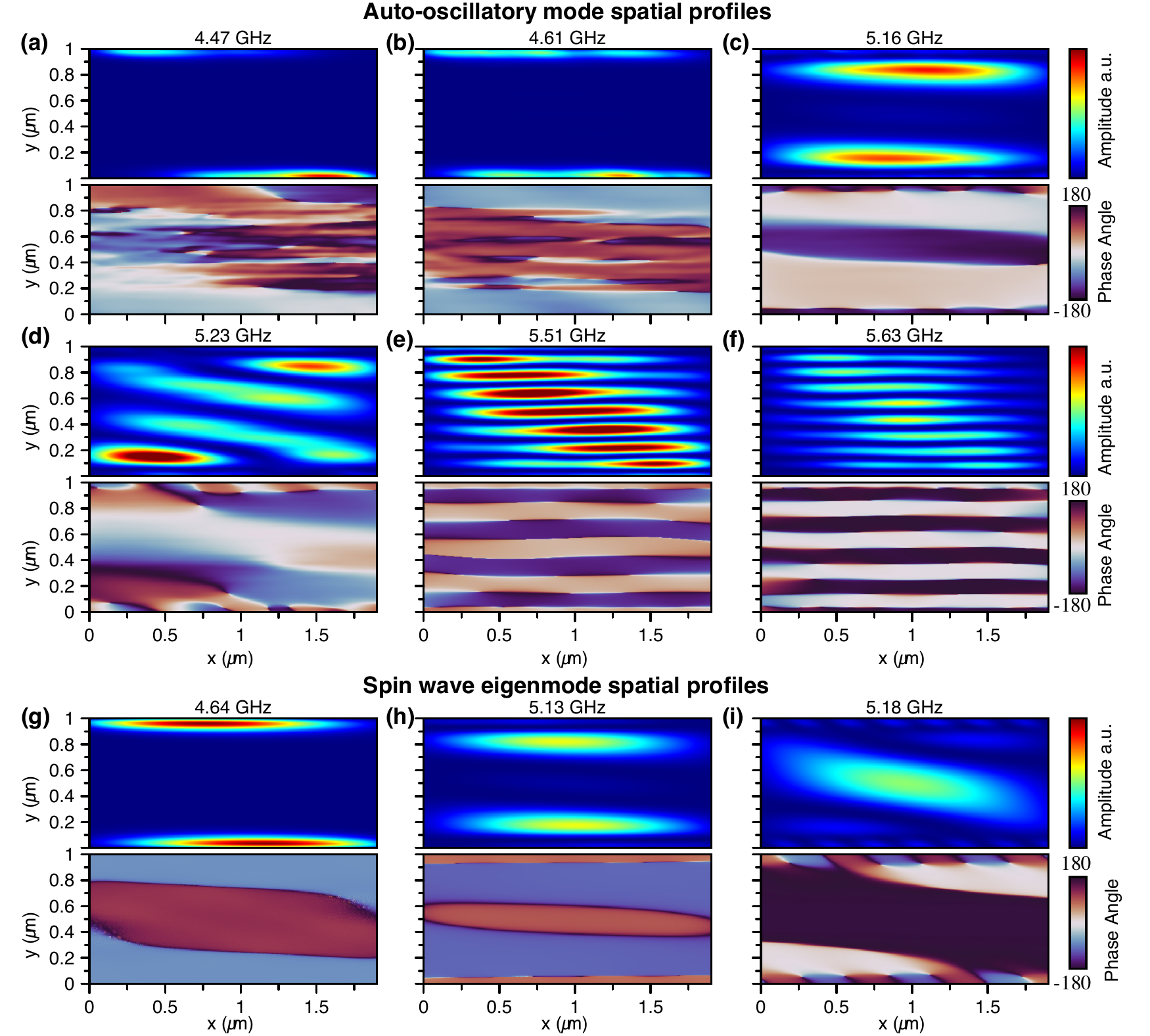}
\caption{\label{fig:Spatial_Profiles}
Micromagnetic simulations of spin wave spatial profiles in the 1.07 $\mu$m wide wire. Amplitude (top) and phase (bottom) spatial profiles are shown for the auto-oscillatory modes with frequencies: (a) 4.47, (b) 4.61, (c) 5.16, (d) 5.23, (e) 5.51, and (f) 5.63. The amplitude scales for (a)--(c), are the same, while the scales for (d), (e), and (f) are $2 \times$, $100 \times$, and $10 \times$ larger respectively. Amplitude (top) and phase (bottom) spatial profiles for spin wave eigenmodes with frequencies: (g) 4.64, (h) 5.13, and (i) 5.18 GHz}. Individual panel labels correspond to the peaks labeled $a$--$i$ in Fig.~\ref{fig:Spectra}.
\end{figure*}

Figs.~\ref{fig:Spatial_Profiles}(a)--\ref{fig:Spatial_Profiles}(f) show the amplitude and phase spatial profiles of the modes labeled $a$--$d$ in Fig.~\ref{fig:Spectra}(a) and  $e$--$f$ in Fig.~\ref{fig:Spectra}(b). The amplitude spatial profiles of the modes lack mirror symmetry due to the 5$^{\circ}$ misalignment of the applied field with the in-plane normal to the wire. Notably, all these modes are localized within the active region of the wire. For the peak cluster just above 4.5 GHz, the spatial profiles reveal the largest amplitude peaks, labeled $a$ and $b$, correspond to edge modes. Spatial profiles reveal that the peaks above 5 GHz, labeled $c$--$f$, correspond to bulk modes \cite{Duan2015}. Note that the amplitude scale is set equal for Figs.~\ref{fig:Spatial_Profiles}(a)--\ref{fig:Spatial_Profiles}(c), while the scale for Figs.~\ref{fig:Spatial_Profiles}(d), \ref{fig:Spatial_Profiles}(e), and \ref{fig:Spatial_Profiles}(f) are $2 \times$, $100 \times$, and $10 \times$ larger respectively. Owing to the large FFT amplitude occupying a significant area of the active region, one might expect modes $e$ and $f$ to have a larger amplitude relative to peaks $a$-$d$ in the emission spectrum shown in Fig.~\ref{fig:Spectra}(a). However, as shown in the corresponding phase profiles, alternating anti-nodes have dynamics that are out of phase which significantly reduces contribution to the global average of the dynamic magnetization. The simulated $\langle m_x^2 \rangle$ spectrum, being insensitive to phase, does indeed show that the largest amplitude peak is mode $e$, which is followed in amplitude by mode $f$.

In order to understand how the auto-oscillatory modes are related to spin wave eigenmodes of the nanowire, we use micromagnetic simulations to calculate the spin wave eigenmode spectrum and spatial profiles. In these simulations, we apply the same Oersted field in the active region, but the spin Hall torque is turned off, effectively keeping the system the same but with no anti-damping torque. An out-of-plane magnetic field pulse of small amplitude $h_\textrm{sinc} = 5~\textrm{Oe}$ and having a time-dependent profile described by the sine cardinal function $h_\textrm{sinc} \textrm{sinc}(t) = h_\textrm{sinc} \sin(2 \pi f_\mathrm{c} t)/(2 \pi f_\mathrm{c} t)$ is applied to the wire. As a result, all spin wave eigenmodes of the wire with frequencies below the cutoff frequency $f_\mathrm{c}=20$\,GHz are excited by the pulse \cite{Dvornik2011,Kumar2017}. We then perform FFT of the $x$-component of the dynamic magnetization $m_x \left( t \right)$ to obtain the spectrum of spin wave eigenmodes shown in Fig.~\ref{fig:Spectra}(c).

Fig.~\ref{fig:Spectra}(c) reveals a similar grouping of spin wave eigenmodes into two clusters at similar frequencies as the auto-oscillatory modes shown in Fig.~\ref{fig:Spectra}(a). The amplitude spatial profiles of the largest amplitude peaks $g$--$i$ are shown in Figs.~\ref{fig:Spatial_Profiles}(g)--\ref{fig:Spatial_Profiles}(i). The near 4.5 GHz edge mode profiles of auto-oscillatory modes $a$ and $b$ are similar to that of spin wave eigenmode $g$. We observe a similar match with the 5 GHz auto-oscillatory bulk mode $c$ to that of spin wave eigenmode $h$. However, we do not observe an auto-oscillatory mode that matches with the bulk spin wave eigenmode $i$.

\section{Results and Discussion}

\subsection{Comparison and discussion of microwave emission, BLS, and micromagnetic simulations}

Comparing the spin wave eigenmodes and auto-oscillatory modes predicted by micromagnetic simulations to the experimental data provided by electrically detected microwave emission and Brillouin light scattering, we find both qualitative agreement and apparent discrepancies for the 1.07 $\mu$m wire. In both the micromagnetic simulations and BLS data, we observe edge modes near 4.5 GHz. In both spectra, edge mode amplitudes are smaller than the dominant modes above 5 GHz. In the microwave emission measurements, however, we do not clearly detect modes near 4.5 GHz. As stated before, this discrepancy may be attributed to the fraction of the active region occupied by the edge mode being too small to generate microwave signal above our instrument's noise floor. This is supported by the data for narrower wires, where edge modes occupying a larger volume fraction of the active region are clearly observed in the microwave emission spectra near 4.5 GHz.  Additionally, edge roughness or damage may lead to varying edge anisotropy and/or current distributions that significantly affect the mode profile and excitation, which is not assumed in micromagnetic simulation \cite{McMichael2006}. The amplitude asymmetry of the 4.5 GHz edge mode as detected by BLS (Fig.~\ref{fig:BLS}(c)) indicates that the two edges of the nanowire are in fact not identical.

All the data show large excitation of magnetization dynamics at frequencies above 5 GHz. Electrically detected emission measurements (Fig.~\ref{fig:Emission}(a)) and global-average auto-oscillatory micromagnetic simulations (Fig.~\ref{fig:Spectra}(a)) exhibit a similar grouping of peaks just above 5 GHz. In the BLS experiment (Fig.~\ref{fig:BLS}(b)) one peak centered around 5.6 GHz is detected, which is expected from the cell-by-cell average micromagnetic simulations (Fig.~\ref{fig:Spectra}(b)). Spatial profiles from both micromagnetic simulations and BLS indicate these modes above 5 GHz are bulk modes. 

The micromagnetic simulations indicate that the bulk mode making dominant contribution to the microwave emission exhibits maximum amplitude relatively far from the middle of the wire, while the BLS data indicate that the majority of the bulk of the nanowire is excited by spin Hall torque. To understand this apparent discrepancy, we must consider that BLS is sensitive to the total population of magnons. We find that higher-order bulk modes, which lead to small-to-negligible amplitude in microwave emission due to phase cancellations, can lead to large amplitude in BLS spectra. Simulations predict that higher frequency modes should actually dominate the BLS spectra compared to the emission spectra, see Figs.~\ref{fig:Spectra}(a) and \ref{fig:Spectra}(b). In fact this is what we observe experimentally; the BLS spectrum in Fig.~\ref{fig:BLS}(b) appears shifted to higher frequency compared to the emission spectrum in Fig.~\ref{fig:Emission}(a). 

We note that the microwave emission and BLS experiments were performed with slightly different external fields (470 Oe for emission and 495 Oe for BLS) and at different temperatures (4.2 K for emission and 300 K for BLS). Therefore care must be taken when comparing mode frequencies between the two experiments. As Py is a low anisotropy material, only two temperature dependent magnetic parameters influence the frequency of the modes: saturation magnetization $M_\mathrm{s}$ and exchange stiffness constant $ A_\mathrm{ex} $. Assuming $A_\mathrm{ex} \propto M_\mathrm{s}^2$ and a decrease in $M_\mathrm{s}$ of 7\% upon increasing temperature from 4.2 K to 300 K \cite{Luo2015}, our micromagnetic simulations predict a frequency shift of $\Delta f \approx -0.20$ GHz for bulk eigenmodes near 5.5 GHz.  The  25 Oe increase in applied field for the BLS measurements shifts the bulk eigenmode frequency by $\Delta f \approx +0.15$ GHz. Therefore the temperature-induced decrease of mode frequency is nearly compensated by the field-induced increase of frequency for BLS measurements. Thus we expect the frequencies of the modes measured in our room temperature BLS experiment in Fig. \ref{fig:BLS}(b) to be similar to those measured in the microwave emission measurements at 4.2 K in Fig. \ref{fig:Emission}(a).

The micromagnetic simulations yield reasonable insight into the observed discrepancies between the spectra for microwave emission and BLS experiments. We find that the two experiments have different sensitivity to different auto-oscillatory modes. The microwave emission measurement is sensitive to phase cancellations and we find that in-phase bulk modes localized away from the wire center are the dominant source of the generated microwave signal. BLS is sensitive to the number of excited magnons and, therefore, bulk modes excited with large amplitude dominate the BLS signal. 

\begin{figure*}[ht]
\includegraphics{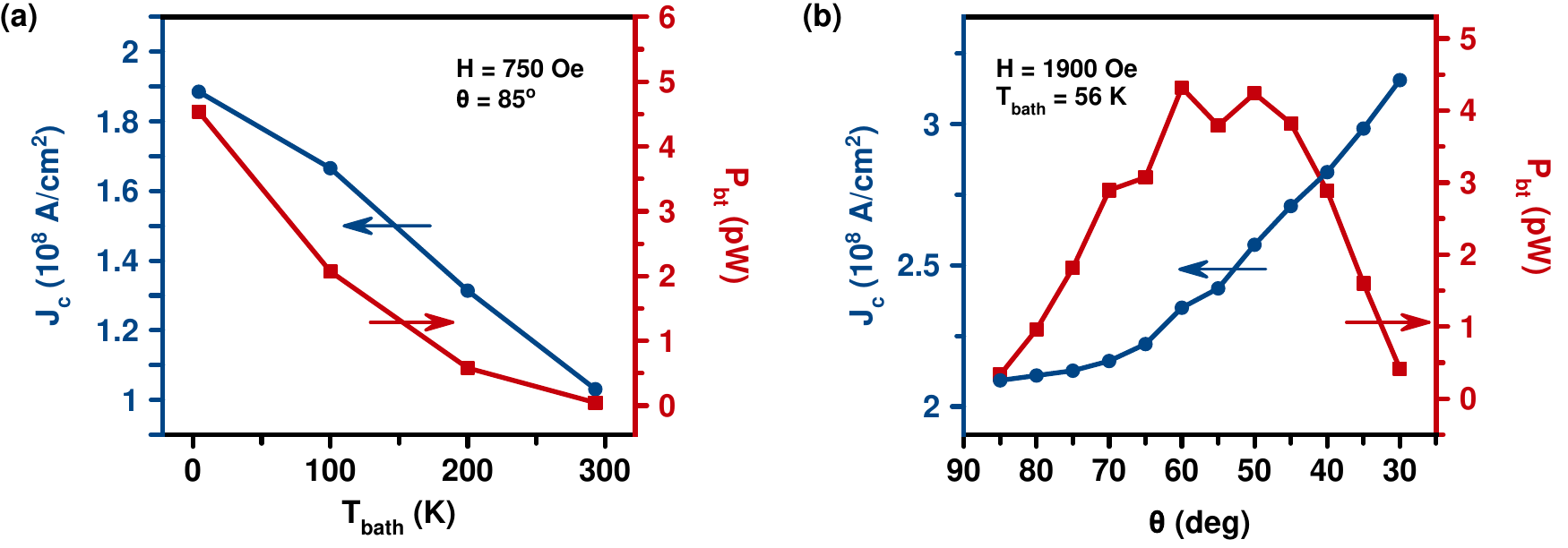}
\caption{\label{fig:Temp_Angle_Dep}
Temperature and angular dependence of auto-oscillatory dynamics in the 0.53 $\mu$m wide wire device. 
(a) Critical current density $J_\mathrm{c}$ (blue circles) and integrated power $P_\mathrm{bt}$ (red squares) versus bath temperature measured at $H=750$ Oe and $\theta = 85^\circ$. (b) Critical current density (blue circles) and maximum emitted power  $P_\mathrm{bt}$ (red squares) versus in-plane magnetic field angle $\theta$ measured at bath temperature of 56 K and magnetic field $H =1900$ Oe.}
\end{figure*}

It may come as a surprise that the higher order bulk modes $e$ and $f$ are excited with large amplitude while lower order bulk modes are not. We propose an explanation of this effect via nonlinear mode damping. In our measurements, the external field is applied nearly perpendicular to the nanowire. Therefore, the observed bulk modes arise from geometric confinement of backward volume dipole-exchange modes in a 2D film  \cite{Gurevich1996}, for which the longest wavelength modes do not have the lowest frequency. In fact, we find that the first few bulk modes are closer to each other in frequency, see inset of Fig.~\ref{fig:Spectra}(b), consistent with a local minima of the dispersion relation at a non-zero wave vector expected for backward volume spin waves. The near degeneracy of these modes enhances coupling among them and thus increases nonlinear damping of each mode \cite{Melkov2013,Slobodianiuk2019,Barsukov2019}. Enhanced nonlinear damping impedes excitation of the auto-oscillations and limits their amplitude \cite{Slavin2009}. On the other hand, higher order modes have weaker nonlinear damping due to more sparse mode spectrum at higher frequencies; thus, these modes may be easier to excite into the auto-oscillatory regime. A quantitative theory will be necessary to fully explain this effect observed in the numerical simulations.

It is instructive to estimate the characteristic wire width of the 1D to 2D dimensional crossover in our experiments. A  rough estimate of the crossover width can be obtained from the backward volume spin wave dispersion relation for a given thickness of the ferromagnetic film \cite{Gurevich1996}. Due to the non-monotonic dependence of the backward volume spin wave frequency on wave number, the uniform $\mathbf{k=}0$ mode can undergo energy- and momentum-conserving four-magnon scattering into backward volume spin wave modes with a non-zero wave vector $ \mathbf{k}_4$ \cite{Gurevich1996}. In the wire geometry, the backward volume spin wave mode spectrum becomes quantized and the four-magnon scattering channel is suppressed when the width is reduced below $ \sim \pi k_4^{-1} $ \cite{Duan2014}.  The suppression of such nonlinear magnon scattering processes allows for large amplitude auto-oscillatory modes to be excited and sustained above the critical current. In fact, \citet{Duan2014} calculated the dispersion for 5 nm thick Py, like that used in the present study, and found that $ \pi k_4^{-1}  \approx 0.5 \, \mu \mathrm{m} $. This is consistent with our data in  Fig.\,\ref{fig:Emission} and Fig.\,\ref{fig:Emission_Power} showing a strong decrease in the emitted microwave power from bulk modes and multi-mode auto-oscillatory dynamics for wires significantly wider than 0.5 $\mu$m. 

As a guiding principle in the design of wire spin torque oscillators, one should aim to suppress the nonlinear magnon scattering channels that lead to enhanced nonlinear damping. Our work demonstrates that reduction of the wire width below a characteristic crossover width is a viable approach to this task. It is clear that the characteristic crossover width increases with increasing exchange and decreases with increasing saturation magnetization. Another interesting approach leading to suppression of nonlinear damping has been recently reported by \citet{Divinskiy2019}. In this approach, nonlinear mode coupling and nonlinear damping are suppressed via cancellation of magnetic shape anisotropy in a ferromagnetic film by perpendicular magnetic anisotropy. This cancellation strongly reduces spin wave ellipticity, which results in weak nonlinear damping \cite{Demidov2020}.

\subsection{\label{sec:temp}Angular and temperature dependence of microwave emission}

In order to better understand the factors that determine spectral properties of the microwave signal generated by the SHOs, we make measurements of the emitted microwave signal as a function of bath temperature. Fig.\,\ref{fig:Temp_Angle_Dep}(a) shows the dependence of the total microwave power emitted by all bulk modes $P_\mathrm{bt}$ of the 0.53 $\mu$m wide wire device versus bath temperature measured at $H=750$\,Oe and $\theta=85^\circ$. The generated power monotonically decreases with increasing bath temperature, dropping by nearly two orders of magnitude between 4.2\,K and 295\,K. At the same time, the critical current density for the onset of auto-oscillatory dynamics $J_\mathrm{c}$ decreases by approximately a factor of two. 

The precipitous decrease of the coherent microwave signal generated by the SHO with increasing temperature points to the important role played by incoherent thermal magnons in limiting the amplitude of coherent auto-oscillatory dynamics. Since the rate of non-linear magnon-magnon scattering processes generally increases with increasing magnon population \cite{Suhl1957},  the rate of scattering from large-amplitude coherent auto-oscillatory modes into incoherent magnon modes increases with increasing thermal population of the incoherent modes. This leads to a significant decrease of the amplitude of the auto-oscillatory modes as observed in our measurements. In the confined geometry studied in present work, both resonant and non-resonant nonlinear magnon scattering processes are important \cite{Melkov2013,Slobodianiuk2019,Barsukov2019}.  A detailed theoretical model is required for a quantitative explanation of the observed strong temperature dependence of the amplitude of auto-oscillations.

The generated microwave power $P_\mathrm{bt}$ and the critical current density $J_\mathrm{c}$  depend not only on the temperature of the ferromagnet, but also on the magnitude and direction of the applied magnetic field $H$. In agreement with the general theory of spin torque oscillators \cite{Slavin2009}, we find the critical current density to increase with increasing magnitude of in-plane magnetic field as well as with increasing angle between magnetization and spin Hall current polarization $ \left| 90^\circ - \theta \right|$. 

Fig.~\ref{fig:Temp_Angle_Dep}(b) shows the measured dependence of $J_\mathrm{c}$ and $P_\mathrm{bt}$ on the in-plane applied magnetic field direction $\theta$ for the 0.53 $\mu$m wide wire device. For these measurements, we employ a 1.9\,kOe magnetic field that is large enough to ensure collinearity of the magnetization with the applied field direction. The critical current due to the antidamping spin Hall torque is expected to increase with angle between magnetization and spin Hall current polarization as $J_\mathrm{c0}/\cos( 90^\circ - \theta )$, where $J_\mathrm{c0}$ is the critical current density for magnetization parallel to the spin Hall current polarization. The data in Fig.~\ref{fig:Temp_Angle_Dep}(b) are qualitatively consistent with this expression. 

The angular dependence of the generated microwave power $P_\mathrm{bt}(\theta)$ shown in Fig.~\ref{fig:Temp_Angle_Dep}(b) is governed by two factors: (i) the angular dependence of AMR: $R = R_\mathrm{min}+\Delta R_\mathrm{AMR} \cos^2(\theta)$, which results in the maximum efficiency of conversion of magnetization oscillations into resistance oscillations for magnetization making a 45$^\circ$ angle with respect to the electric current direction and (ii) the angular dependence of the critical current density $J_\mathrm{c}(\theta)$. The data for $P_\mathrm{bt}(\theta)$ in Fig.~\ref{fig:Temp_Angle_Dep}(b) exhibit a maximum for $\theta$ between 60$^\circ$  and 50$^\circ$.  This angle of the maximum emitted power deviates from 45$^\circ$ because of the angular dependence of the critical current density. Indeed, rotation of the magnetization away from the $y-$axis $\left( \theta=90^\circ \right)$ increases the critical current, as shown by the blue circles in Fig.~\ref{fig:Temp_Angle_Dep}(b), which may result in a decrease of the emitted power at a fixed value of $J_\mathrm{dc}$.

\section{\label{sec:results}Summary}

This work provides a systematic experimental investigation of the 1D to 2D dimensional crossover of current-driven auto-oscillatory dynamics in spin torque oscillators. We study such auto-oscillatory dynamics in a series of Pt/Py wire-based spin Hall oscillator devices with varying wire width. Our measurements reveal graduate disappearance of coherent auto-oscillatory dynamics with increasing wire width. This crossover proceeds via increase of the number of auto-oscillatory modes with increasing wire width in conjunction with a rapid decrease of the maximum amplitude of auto-oscillations reached by each of the modes. This decrease of the amplitude of auto-oscillatory modes can be explained by mode competition for the total angular momentum provided by the spin Hall current and by nonlinear interactions among spin wave modes, which limit the amplitude of each of the modes. Temperature-dependent measurements reveal a precipitous decrease of the amplitude of auto-oscillatory dynamics with increasing sample temperature. We ascribe this decrease to nonlinear scattering of coherent auto-oscillatory modes on incoherent thermal magnons. The amplitude of this nonlinear scattering process increases with temperature due to an increase of the incoherent magnon population. 

\bigskip

\section*{Acknowledgments}

This work was supported by the National Science Foundation through Grants No.\,DMR-1610146, No.\,EFMA-1641989 and No.\,ECCS-1708885. We also acknowledge support by the  Army Research Office through Grant No.\,W911NF-16-1-0472, Defense Threat Reduction Agency through Grant No.\,HDTRA1-16-1-0025 and the Beall Innovation Award. Brillouin light scattering measurements were supported through Spins and Heat in Nanoscale Electronic Systems (SHINES), an Energy Frontier Research Centre funded by the US Department of Energy, Office of Basic Energy Sciences under Award no. DE-SC0012670

\end{document}